\documentclass[12pt,preprint]{aastex}

\begin{document}


\title{Chemical Abundances in the Secondary Star\\ in the Black Hole Binary \mbox{A0620$-$00}}

\author{Jonay I. Gonz\'alez Hern\'andez, Rafael Rebolo,\altaffilmark{1}\\ Garik
Israelian, and Jorge Casares}

\affil{Instituto de Astrof{\'\i }sica de Canarias, E-38205 La Laguna, Tenerife,
SPAIN: \\jonay@ll.iac.es, rrl@ll.iac.es, gil@ll.iac.es, jcv@ll.iac.es}

\and

\author{Andre Maeder and Georges Meynet}

\affil{Geneva Observatory, 1290 Sauverny,
SWITZERLAND: \\Andre.Maeder@obs.unige.ch, Georges.Meynet@obs.unige.ch}

\altaffiltext{1}{Consejo Superior de Investigaciones Cient{\'\i }ficas, SPAIN}

\begin{abstract}
Using a high resolution spectrum of the secondary star in the black hole binary
\mbox{A0620$-$00}, we have derived the stellar parameters and  veiling caused by
the accretion disk in a consistent way. We have used a $\chi^{2}$ minimization
procedure to explore a grid of 800\,000 LTE synthetic spectra computed for a
plausible range of both stellar and veiling parameters. Adopting the best model
parameters found, we have determined atmospheric abundances of Fe, Ca, Ti, Ni
and Al. The Fe abundance of the star is $\mathrm{[Fe/H]}=0.14 \pm 0.20$. Except for Ca, we found the other elements
moderately over-abundant as compared with stars in the solar neighborhood of
similar iron content. Taking into account the small orbital separation, the mass
transfer rate and the mass of the convection zone of the secondary star, a
comparison with element yields in supernova explosion models suggests a
possible explosive event with a mass cut comparable to the current mass of the
compact object. We have also analyzed the Li abundance, which is unusually high for a
star of this spectral type and relatively low mass.   
\end{abstract}

\keywords{black holes physics---stars:abundances---stars:individual
(\mbox{A0620-00})---stars:X-rays:low-mass---binaries}

\section{Introduction}

The system \mbox{A0620$-$00} (V616 Mon) is a low mass X-ray binary (LMXB)
discovered as an eruptive X-ray source by {\it Ariel V} in August 1975 (Elvis 
et al., 1975). During the outburst, it brighte\-ned by 6 magnitudes in the
optical and after 15 months it had returned to its quiescent magnitude of 
$m_{V}=18.3$ mag. Spectroscopic observations during quiescence revealed a K5\,V--K7\,V
stellar spectrum plus an emission line component from an accretion disk
surrounding the compact object (Oke 1977; Murdin et al., 1980). Further optical
photometric and spectroscopic studies established the orbital period at $P =
0.323$ d and a secondary radial velocity semiamplitude of $K_2 = 457$ {${\rm km}\:{\rm s}^{-1}$}
(McClintock \& Remillard 1986), which implied a mass function of $f(M) = 3.18 \pm
0.16$ {$M_\odot$} and thus firm dynamical evidence for a massive compact 
object---a black hole---in this system. From measurements of the orbital inclination, the
compact object mass was estimated at $\sim 11$ {$M_\odot\; $} and the companion star mass
at $\sim 0.7$ {$M_\odot\; $} (Shahbaz et al., 1994; Gelino et al., 2001). 

Many aspects of the origin and evolution of low mass X-ray binaries (LMXBs)
still remain unclear. It is believed that these systems begin as wide binaries
with extreme mass ratios and orbital separations of $a
\sim 1000$ {$R_\odot\; $} (Portegies Zwart et al., 1997; Kalogera \& Webbink 1998). After
filling its Roche lobe, the massive star engulfs its low mass companion and the
latter starts to spiraling in to the massive star's envelope (van den Heuvel \&
Habets 1984; de Kool et al., 1987). A close binary forms if the spiral-in ceases
before the low mass companion coalesces with the compact helium core of the
primary. The helium core continues its evolution and after SN explosion may turn
into a neutron star or a black hole. The system becomes an X-ray binary once the
secondary star fills its Roche lobe and begins to transfer matter to the compact object.

The spiral-in process could give rise to a naked He core, identified with
Wolf--Rayet stars that have lost their envelopes (Woosley et al., 1995). The high
mass-loss rate (Chiosi \& Maeder 1986; Nugis \& Lamers 2000) of these stars makes difficult to
understand the formation of compact objects as massive as the black hole in
\mbox{A0620$-$00} (Meynet \& Maeder 2003; Woosley et al., 1993). However, if the hydrogen envelope of the 
massive star is removed at the end of the He core burning phase (the so-called
{\it Case C} mass transfer, Brown et al., 1999), the mass lost by wind in the
short-lived ($\sim 10^4$ yr) supergiant stage will not be large.  

Convection (Langer 1991) and rotation (Maeder \& Meynet 2000; Heger et al., 2000)
influence the structure and evolution of massive stars and subsequently the
uncertainties in the treatment of these parameters limit our understanding of
the evolution of the progenitors of compact objects. In addition, uncertainties
in various aspects of the supernova explosion models affect the predictions of
the final remnant mass and the chemical composition of any ejecta captured
by the companion. Among the  least known ingredients of these models, we may
list: 

\begin{itemize}

\item The {\it mass cut}, i.e. the mass above which the matter is expelled at the
time of the supernova explosion and below which it remains locked into the
compact remnant. 

\item The {\it amount of fallback} or of the mass which is eventually accreted by
the compact core (Woosley \& Weaver 1995; MacFadyen et al., 2001).  

\item Possible {\it mixing} during the collapse phase (Herant \& Woosley 1994;
Herant et al., 1994; Kifonidis et al., 2000; Fryer \& Warren 2002).  

\item The energy of the supernova explosion (Nakamura et al., 2001).  

\item The symmetry of the supernova explosion (MacFadyen \& Woosley 1999; Maeda et al., 2002).  

\end{itemize}

With the aim of obtaining information on the link between compact objects and their
progenitor stars, Israelian et al. (1999) measured element abundances in the
secondary star of the black hole binary Nova Scorpii 1994 (GRO J1655$-$40) and
found several $\alpha$-elements (O, Mg, Si, S, and Ti) enriched by a factor of
6--10. Since these elements cannot be produced in a low mass
se\-con\-da\-ry star, this was interpreted as evidence of a supernova event that
originated the compact object. Taking into account the supernova yields from
explosion models of massive stars, the re\-la\-ti\-ve abundances of these
elements suggested that the supernova progenitor was in the mass range 25--40
{$M_\odot$}. Afterwards, these over-abundances were compared with a variety of
supernova models, including standard as well as hypernova models (for various
helium star masses, explosion energies, and explosion geometries) and a simple
model of the evolution of the binary and the pollution of the secondary (Brown 
et al., 2000; Podsiadlowski et al., 2002). Additional independent evidence for the
existence of a supernova event in this system has also been found by Mirabel
et al. (2002). 

In this paper we analyze the chemical abundances of the secondary star in the
LMXB \mbox{A0620$-$00} with the aim of searching for any evidence of nucleosynthetic
products from the progenitor of the compact object. 

\section{Observations}

\subsection{Stellar Spectrum}

The secondary star was observed with the UV--Visual Echelle Spectrograph (UVES)
at the Very Large Telescope (VLT) of the European Southern Observatory
(Paranal), using a configuration that provided a dispersion 0.029 \AA/pixel in
the blue arm (4800--5800 \AA) and 0.035 \AA/pixel in the red  (5800--6800
\AA). Short exposures (480--540 s) were chosen in order to avoid as far as 
possible the smearing of spectral lines associated with radial velocity
change during its orbital motion. Twenty spectra were obtained during three
nights in December 2000 in both the blue and the red spectral regions. 

Every spectrum was reduced within the MIDAS UVES environment. First, bias and
inter-order backgrounds were subtracted from both the science and flat-field frames.
The spectrum of the target was then optimally extracted and divided by the
flat-field (extracted with the same weighted profile as the star). The final
spectra were wavelength-calibrated, every order was extracted, and all were
merged in order to obtain a one dimensional spectrum for each 
spectrum.  

\begin{figure}[ht!]
  \centering
  \includegraphics[width=11cm,angle=90]{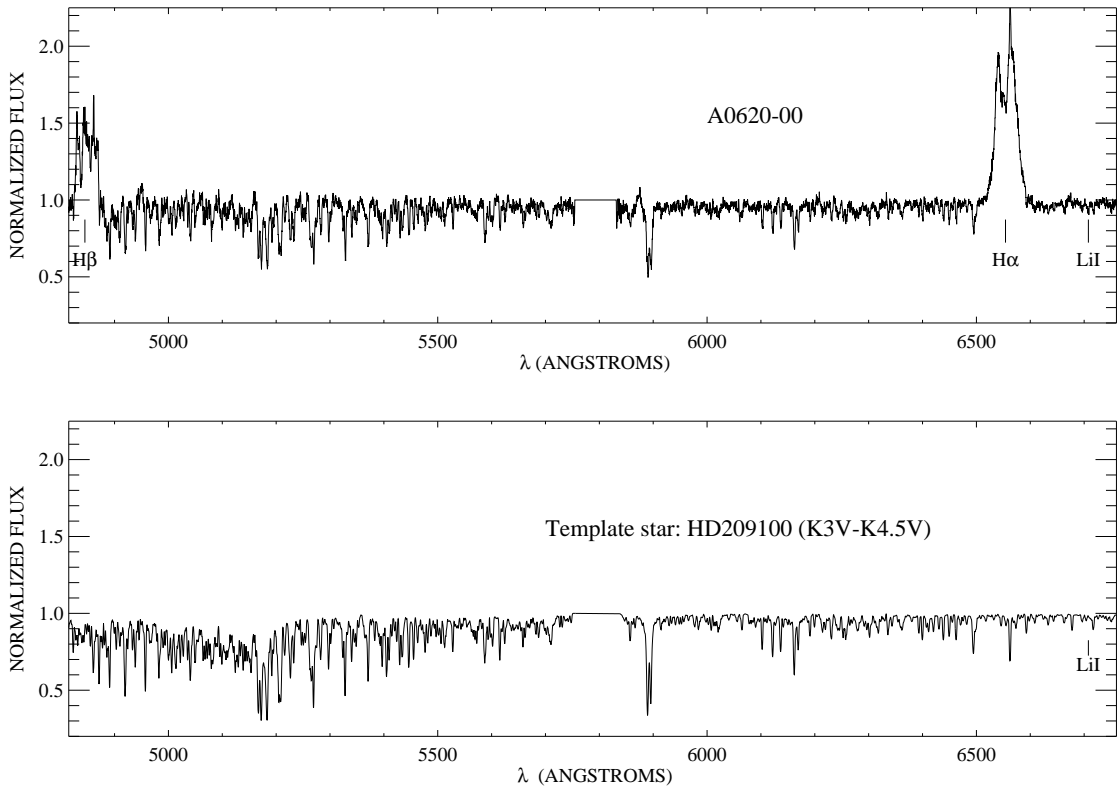}
  \caption{Observed spectrum of the secondary star of \mbox{A0620$-$00} (top)
  and of a properly broadened template (HD209100, bottom). \label{fig1}} 
\end{figure}

The radial velocity for each  spectrum has been obtained from the
ephemeris reported in Casares et al. (2004, in preparation). The individual
spectra were corrected from radial velocity and combined in order to improve
the signal-to-noise ratio. After binning in wavelength in groups of 10 pixels
the final spectrum had a signal-to-noise ratio of 85 in the continuum. This
spectrum is shown in Fig.\ 1.  

\subsection{Telluric Spectrum}

We obtained a telluric spectrum from our own observations and the stellar
spectrum was properly corrected for it. The large difference in broadening
between the stellar  ($\sim 95$ {${\rm km}\:{\rm s}^{-1}$}) and telluric lines 
($\sim 9$ {${\rm km}\:{\rm s}^{-1}$},
the instrumental resolution), allo\-wed us to fit a cubic spline to the stellar
features in the binned spectrum of the star, i.e., the final spectrum displayed
in Fig.\ 1. This fit was sampled to the original dispersion provided by the
spectrograph and subtracted from each of the 20 individual unbinned spectra
ta\-king into account the radial velocity of the star in each case. We used this
dispersion to ensure that the telluric lines would not be smoothed. We obtained
20 residual spectra in which we subtracted the stellar features, and which hence 
contain only telluric and interstellar medium (ISM) features. The combination of
these residual spectra gave a noisy telluric spectrum (spectrum 1) which was
cross-correlated with another telluric spectrum (spectrum 2) obtained from a
fast rotating star. The correlation function of both spectra was centered on zero
velocity. 

In order to correct our target observations, we used the much higher S/N
telluric spectrum 2 scaled down (with the IRAF\footnote{IRAF is distributed by 
National Optical Astronomy Observatory, which is
    operated by the Association of Universities for Research in Astronomy, Inc.,
    under contract with the National Science Foundation.} task {\scshape telluric}) to the
strength of the telluric lines in spectrum 1. This scaled telluric spectrum was
shifted according to the radial velocity of the target during each 
exposure. Then we combined all these shifted spectra to generate a final
telluric spectrum in the rest frame of the companion star (spectrum 3), which
was subtracted from the final spectrum of the target.   

\begin{deluxetable}{ccc}
\tabletypesize{\scriptsize}
\tablecaption{Ranges and steps of model parameters \label{tbl-1}}
\tablewidth{0pt}
\tablehead{
\colhead{Parameter} & \colhead{Range}   & \colhead{Step}   
}
\startdata
     {$T_{\mathrm{eff}}$}   & {4000 K $\to$ 5300 K} & 100 K\\
     {$\log g$}    & {3.5 $\to$ 5}       & 0.1 \\
     {[Fe/H]}      & {$-0.8$ $\to$ 1}    & 0.05 \\
     {$f_{4500}$}  & {0 $\to$ 0.4 }      & 0.05 \\
     {$m_0$}       & {0 $\to$ $-$0.00016}  &-1.778E-05\\
\enddata

\end{deluxetable}

\begin{figure}[ht!]
  \centering
  \includegraphics[width=13cm,angle=0]{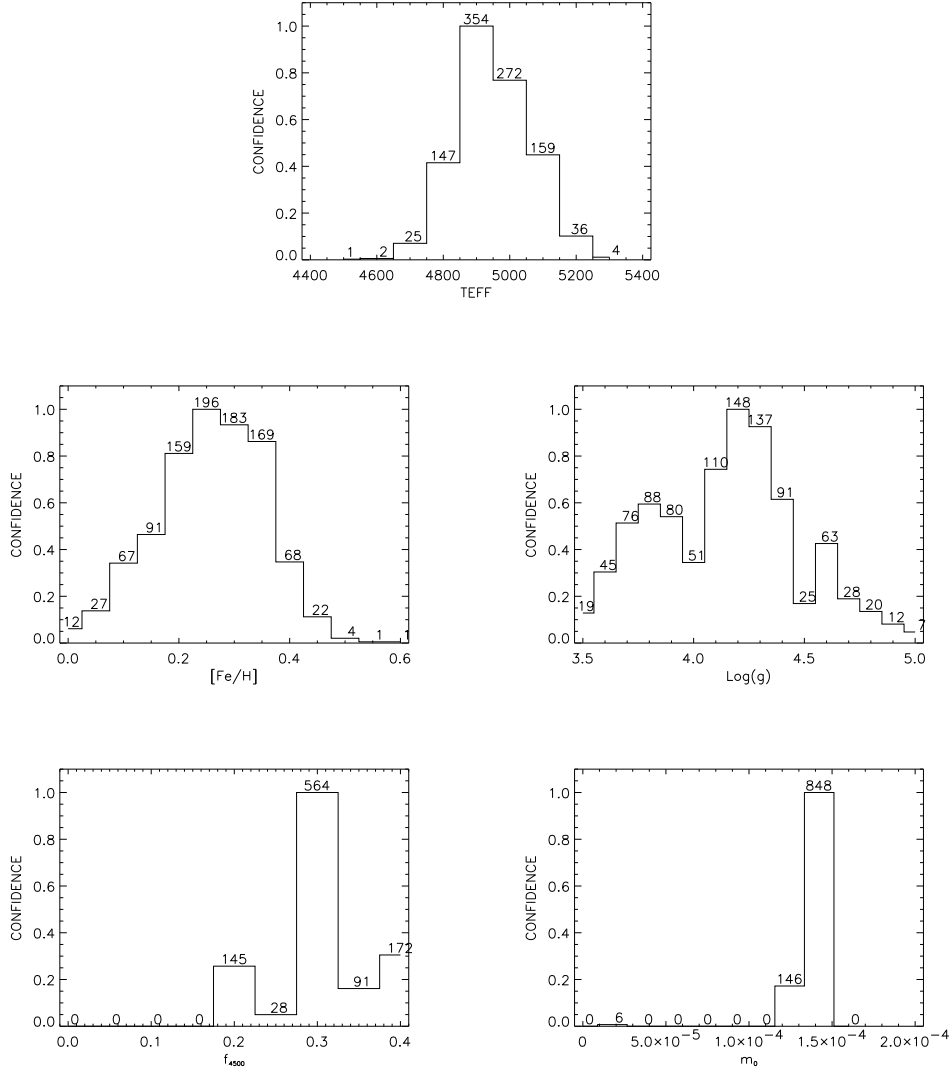}
  \caption{Distributions obtained for each parameter using
  Monte Carlo simulations. The labels at the top of each bin indicate the
  number of simulations consistent with the bin value. The total number of
  simulations was 1000.\label{fig2}}
\end{figure}

\begin{figure}[ht!]
  \centering
  \includegraphics[width=11cm,angle=90]{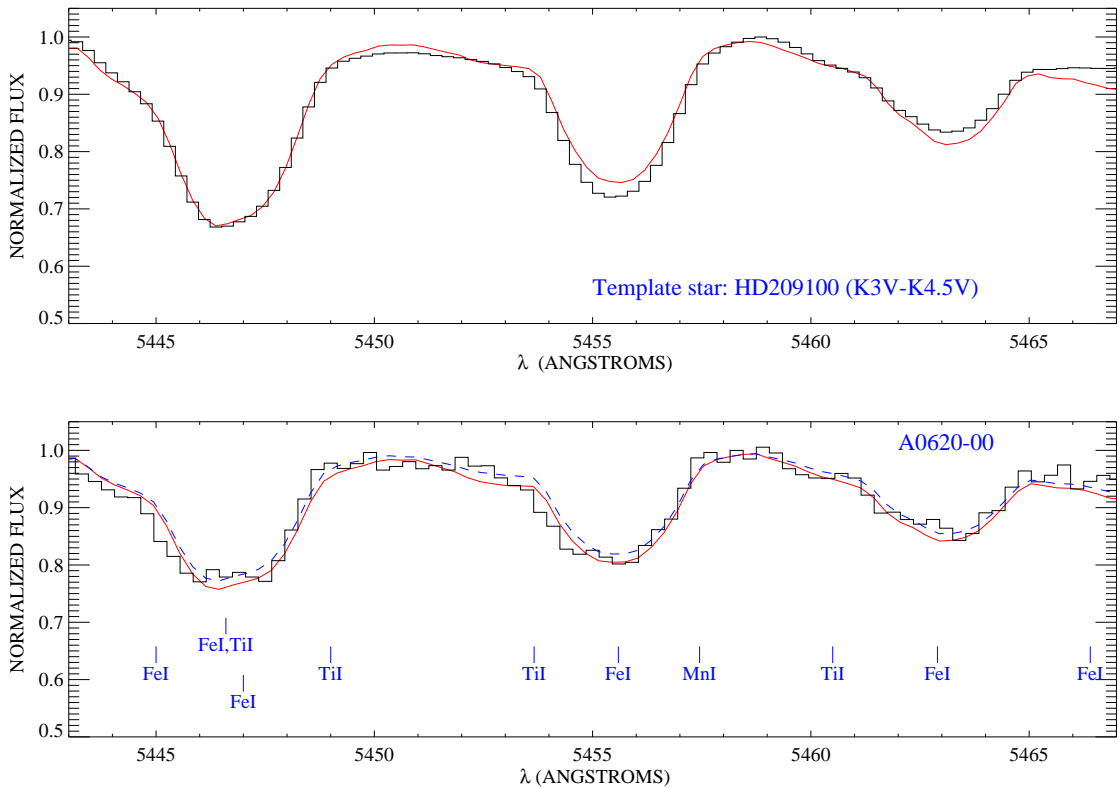}
  \caption{Best synthetic spectral fits to the UVES spectrum of the secondary
    star in the \mbox{A0620--00} system (bottom panel) and the same for a template star
    (properly broadened) shown for comparison (top panel). Synthetic spectra
    are computed for solar abundances (dashed line) and best fit abundance
    (solid line).\label{fig3}}
\end{figure}

\subsection{Diffuse interstellar bands}

In fact, spectrum 1 (see \S 2.2) does not only consist of telluric lines, it
also contains ISM features in the line of sight towards \mbox{A0620-00}. We
tentatively identified in this spectrum several of the well known diffuse
interstellar bands listed in Herbig (1995). We took care that these bands did
not affect the lines selected for the chemical analysis described in the next
section. 

\section{Chemical Analysis}

\subsection{Stellar Parameters}

The chemical analysis of secondary stars in LMXB systems is influenced by three
important factors:  veiling from the accretion disc,  rotational broadening, and 
signal-to-noise ratio. The last two are responsible for the uncertainty in the
continuum position and hence affect the normalization procedure. The veiling
caused by the accretion disk in \mbox{A0620$-$00} appears to drop from the near 
UV ($\sim$30 \%) to the red ($\sim$6\%) (Marsh et al., 1994; McClintock \&
Remillard 2000). However, Shahbaz et al. (1999) have determined that this veiling 
could be as high as 25\% in the IR ($K$ band). All these results depend on
the set of templates used, so we decided to attempt an independent determination
of the veiling. In our analysis we have tried to obtain the veiling, together with
the stellar atmospheric parameters, using synthetic spectral fits to the high
resolution spectrum of the secondary star in \mbox{A0620$-$00}.  
 
First, moderately strong and relatively unblended lines of several elements of
interest were identified in the high resolution solar flux atlas of Kurucz et al. 
(1984). We selected several spectral features containing in total 24 absorption lines of Fe~{\scshape i} with
excitation potentials between 0.5 and 4.5 eV. In order to compute synthetic
spectra for these features, we adopted the atomic line data from the Vienna
Atomic Line Database and used a grid of local thermodynamic equilibrium (LTE)
models of atmospheres provided by Kurucz (1992, private communication). These
models are interpolated for given values of $T_{\mathrm{eff}}$, $\log g$, and
[Fe/H]. Synthetic spectra were then computed \mbox{using} the LTE code MOOG
(Sneden 1973). To mi\-ni\-mi\-ze the effects associated with the errors in the
transition probabilities of atomic lines, we adjusted the oscillator strengths,
$\log gf$ values of the selected lines  until we succeeded in re\-pro\-du\-cing
the solar atlas of Kurucz et al. (1984) with solar abundances (Anders \& Grevesse
1989). 

We generated a grid of synthetic spectra for these features in terms of five
free parameters three to characterize the star atmospheric model (effective
tem\-pe\-ra\-tu\-re, $T_{\mathrm{eff}}$,  surface gravity, $\log g$, and 
metallicity, [Fe/H]) and  two further parameters to take into account the
effect of the accretion disk emission in the stellar spectrum. This veiling was
defined as the ratio of the accretion disk flux to the stellar continuum flux,
$F_{\rm disc}/F_{\rm cont,star}$. It was assumed to be a linear function of
wavelength and is thus characterized by two parameters: veiling at 4500 {\AA},
$f_{4500} = F^{4500}_{\rm disc}/F^{4500}_{\rm cont,star}$, and the slope, $m_0$. These 
five parameters were changed according
to the steps and ranges given in Table 1. Note that the steps for the veiling
slope were chosen to cover all  possible combinations of veiling at different
wavelengths taking into account the step on $f_{4500}$. A rotational
broadening of 95 {${\rm km}\:{\rm s}^{-1}$} and a limb-darkening $\epsilon = 0.65$ were assumed based
on Casares et al. (2004, in preparation), and a fixed value for the
microturbulence, $\xi = 1$ {${\rm km}\:{\rm s}^{-1}$} was adopted.   

The observed spectrum was compared with each of the 800\,000 synthetic spectra in
the grid via a $\chi^2$ minimization procedure that provided the best model fit.
Using a {bootstrap Monte-Carlo method} we defined the 1$\sigma$ confidence
regions for the five free parameters and established as most likely values:
$T_{\mathrm{eff}} = 4900 \pm 100$ K, $\log g = 4.2 \pm 0.3$, [Fe/H]$ = 0.25 \pm 0.1$,
$f_{4500} = 0.30 \pm 0.05$, and $m_0 = -0.00014 \pm 0.00002$.

\begin{deluxetable}{lccccc}
\tabletypesize{\scriptsize}
\tablecaption{Uncertainties in the abundances of the secondary star in
\mbox{A0620$-$00}\label{tbl-3}}   
\tablewidth{0pt}
\tablehead{\colhead{Element} & $\mathrm{[E/H]}_{\rm LTE}$ & \colhead{$\Delta_{\sigma}$}
& \colhead{$\Delta_{T_{\mathrm{eff}}}$} &  \colhead{$\Delta_{\log g}$} &
\colhead{$\Delta_{\rm TOTAL}$\tablenotemark{\dagger}}}
\startdata
Al   &  0.40 & 0.10 & 0.05 & -0.05 & 0.12 \\
Ca   &  0.10 & 0.07 & 0.09 & -0.16 & 0.20 \\
Ti   &  0.37 & 0.18 & 0.10 & -0.05 & 0.23 \\
Fe   &  0.14 & 0.12 & 0.12 & -0.08 & 0.20 \\
Ni   &  0.27 & 0.05 & 0.05 & -0.03 & 0.10 \\
Li\tablenotemark{\star} & 2.31 & 0.10 & 0.15 & 0.01 & 0.21 \\
\enddata
\tablenotetext{\star}{Li abundance is expressed as: \\$$\log \epsilon(\mathrm{Li})_{\rm LTE} =
\log [N(\mathrm{Li})/N(\mathrm{H})]_{\rm LTE} + 12$$}

\tablenotetext{\dagger}{The total error was calculated using the following
formula: \\$$\Delta_{\rm TOTAL} = \sqrt{\Delta_{\sigma}^2 +
\Delta_{T_{\mathrm{eff}}}^2 + \Delta_{\log g}^2 + \Delta_{\rm veiling}^2}$$}  

\tablecomments{The errors from the dispersion of the best fits to different
features, $\Delta_{\sigma}$, are estimated using the following formula: 
$\Delta_{\sigma} =\sigma/\sqrt{N}$ where $\sigma$ is the standard deviation of
the measurements. Total errors also take into account the uncertainties
associated with the stellar parameters and the veiling.}   

\end{deluxetable}

Confidence regions were determined using 1000 realizations. The corresponding
histograms are shown in Fig.\ 2. Our optical veiling determinations are
consistent with previous values in the literature for this system (Marsh et al.,
1994). These authors found a veiling of 17 $\pm$ 3 per cent at H$\beta$ and 6
$\pm$ 3\% at H$\alpha$, while we have obtained (using  their   
definition for veiling), 20 $\pm$ 5 per cent and 0 $\pm$ 5 per cent, respectively.  

\subsection{Stellar Abundances}

Using the derived stellar parameters we ana\-ly\-zed several spectral regions
where we had identified various lines of Fe, Ca, Al, Ti, Ni, and Li. Although
the lines of these elements were usually the main contributor to the features,
in some cases, they were blended, mainly with Fe. The inaccuracy in the
location of the continuum caused by the blends of many weak rotationally
broadened stellar lines was one of the main sources of error in the abundance
determinations. Therefore, each of these spectral regions was carefully
normalized using a late type star template (HD 209100) for comparison.     

We determined the abundances of these elements using spectral synthesis. By
comparing the observed spectrum with a grid of synthetic spectra we identified
the best fit abundance for each element through a $\chi^2$ minimization
procedure. For these spectral syntheses we modified element abundances while
stellar parameters and the suitable veiling factor for each spectral region were
kept fixed. A preliminary estimate of Fe abundance was obtained in the procedure 
described above. We then performed a more detailed analysis of the seven Fe
dominated spectral features but now taking into account the contribution of much
weaker lines to the blends. We obtained an average Fe abundance of
$\mathrm{[Fe/H]}=0.14 \pm 0.12$, where the error is estimated from the dispersion of the abundances inferred from each 
feature. A new model with this metallicity was generated in order to perform
a detailed spectral synthesis for all the features under consideration.

\begin{figure}[ht!]
  \centering
  \includegraphics[width=11cm,angle=90]{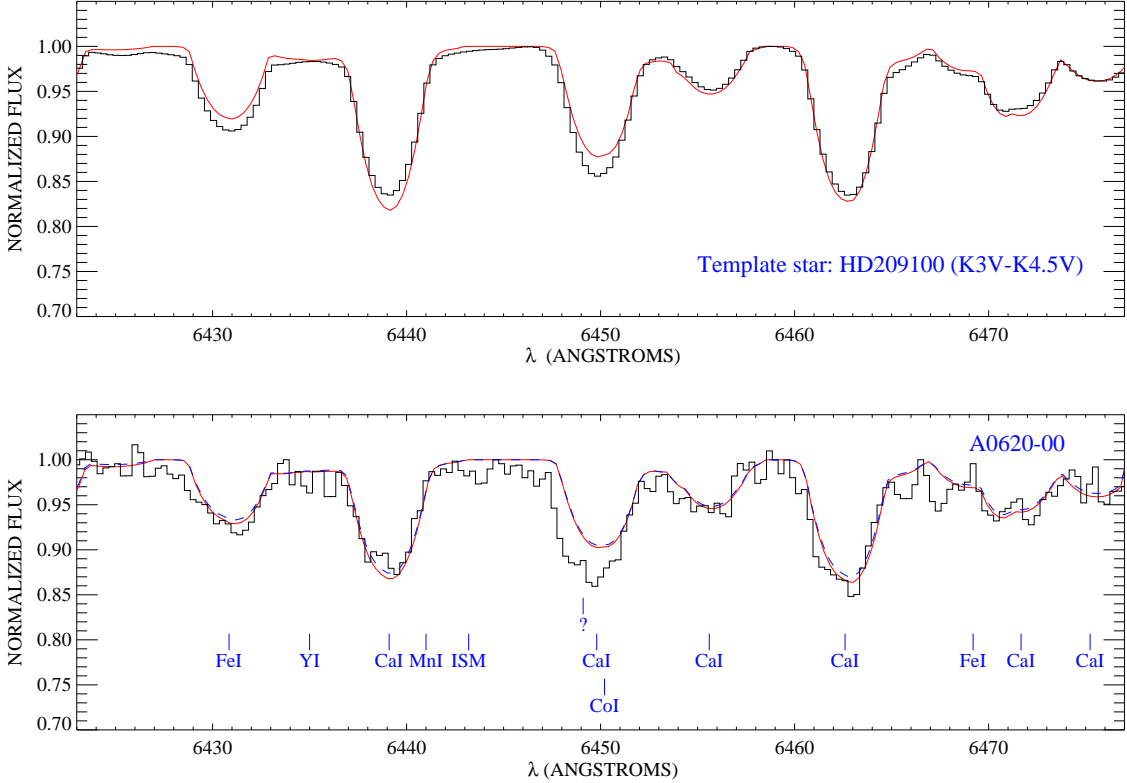}
  \caption{The same as in Fig.\ 3. ``?'' indicates that  
  there is an unidentified line in the solar spectrum. The label ISM
  indicates an identified feature produced by the interstellar medium.\label{fig4}}
\end{figure}

\begin{figure}[ht!]
\centering
\includegraphics[width=5.7cm,angle=90]{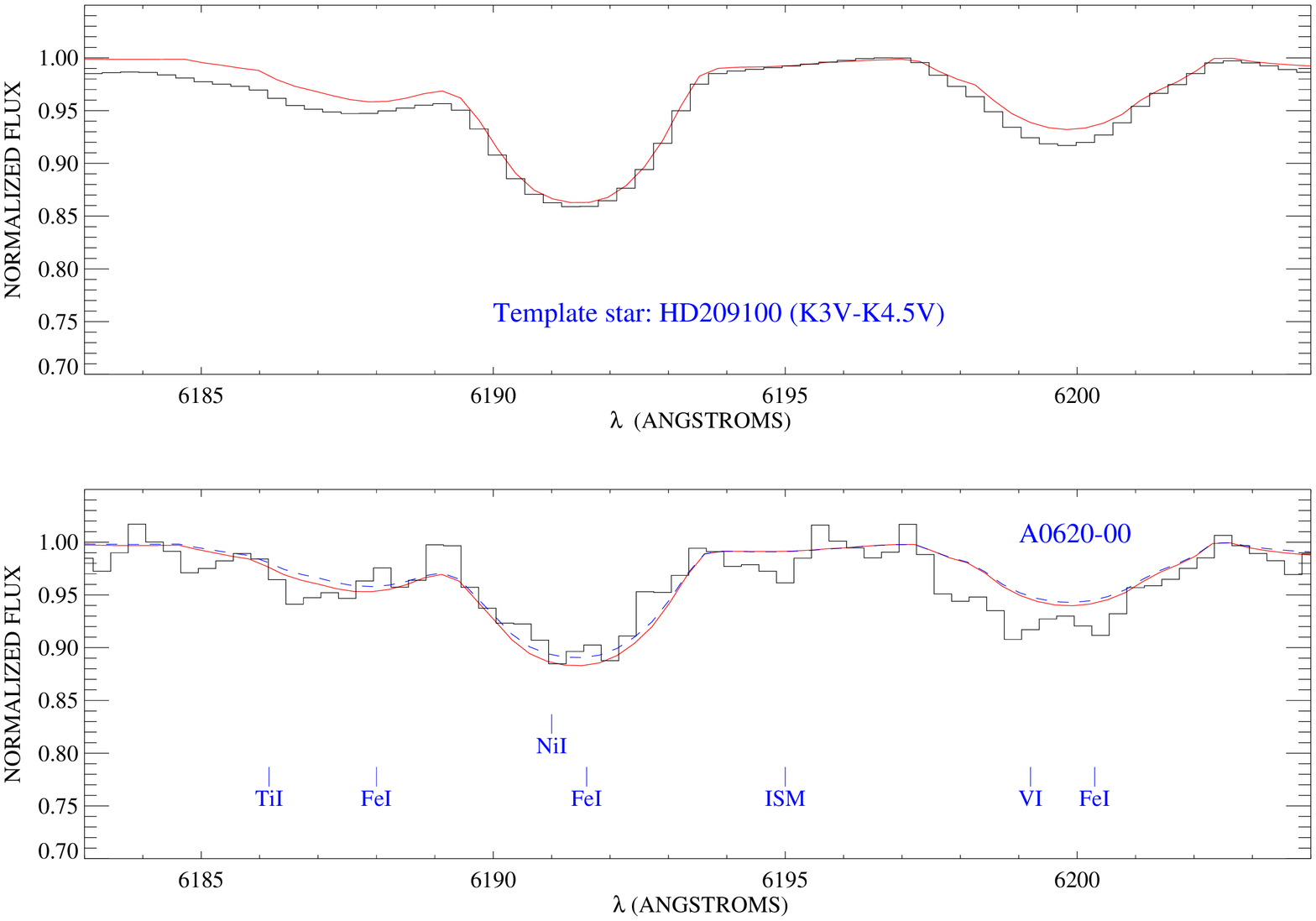}
\includegraphics[width=5.7cm,angle=90]{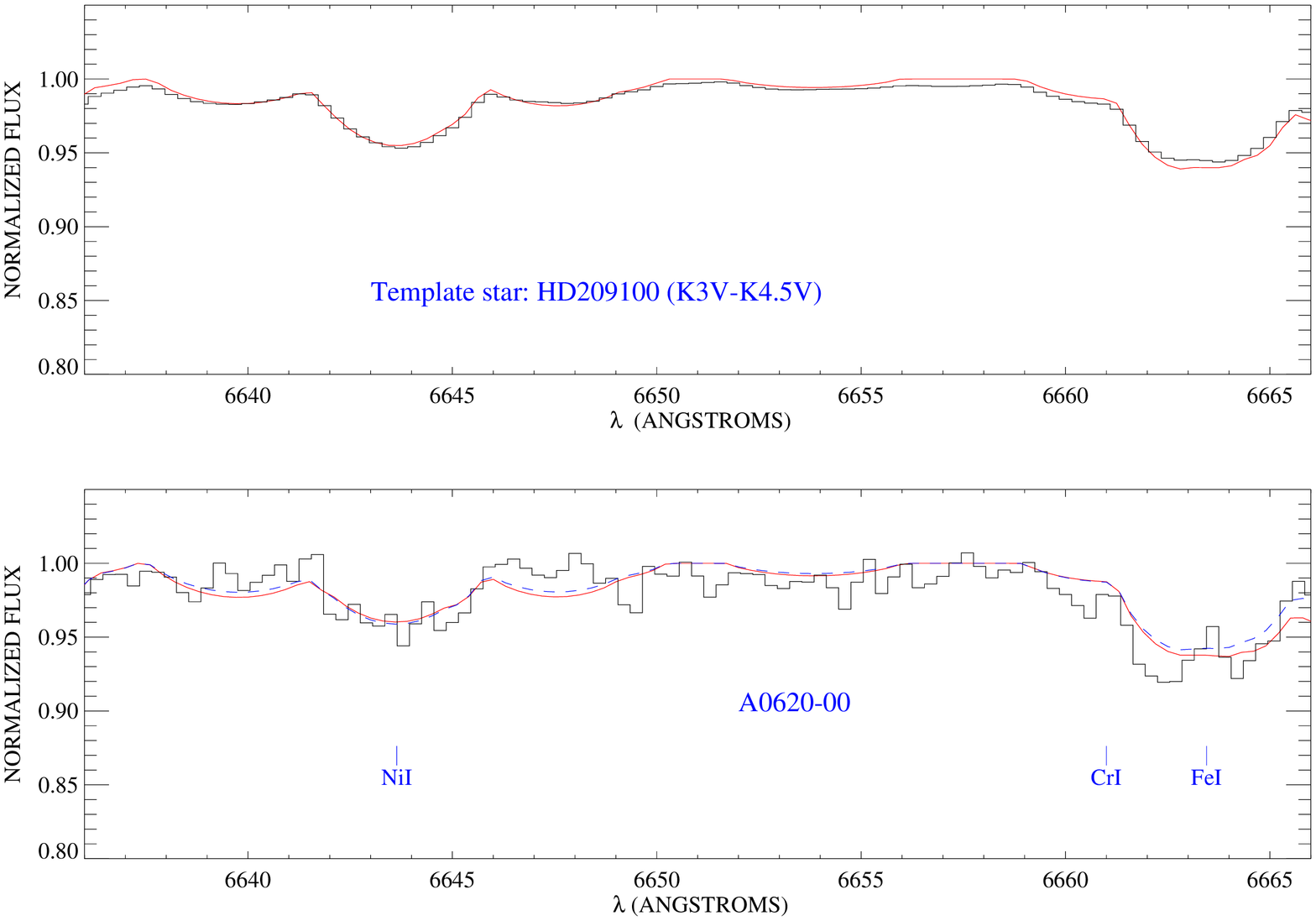}
\protect\caption[ ]{The same as in Fig.\ 3. The Ni line at 6643 {\AA} is
one of the very few isolated lines in the spectrum of the secondary star in
\mbox{A0620$-$00}. The label ISM indicates an identified feature produced by
the interstellar medium.\label{fig5}} 
\end{figure}

\begin{figure}[ht!]
  \centering
  \includegraphics[width=11cm,angle=90]{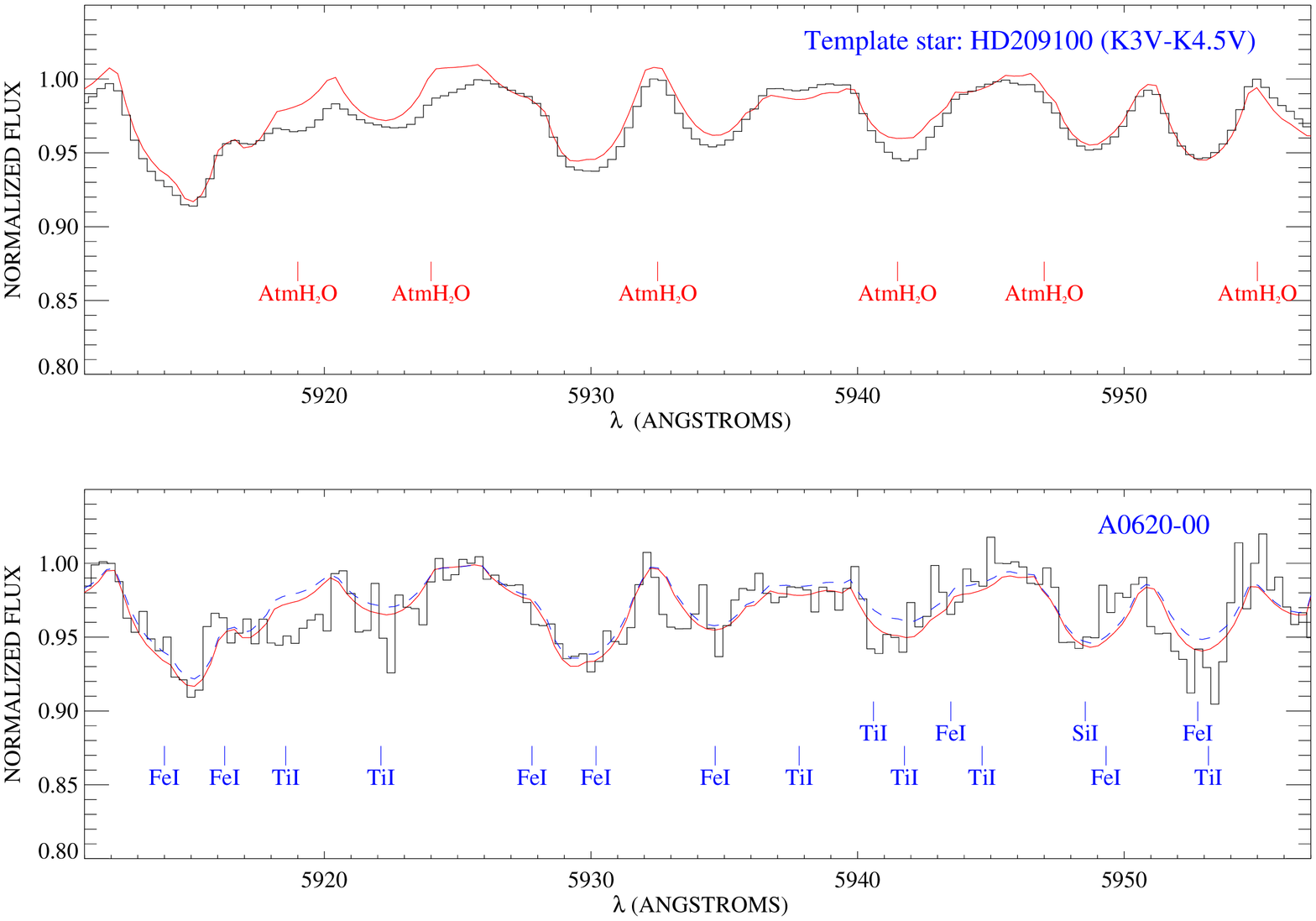}
  \caption{The same as in Fig.\ 3. The spectrum of the template is not
  corrected for telluric lines (atm.\ H$_2$O);  therefore, these lines appear broadened in
  the spectrum (histogram line) and the synthetic spectrum does not fit (solid
  line).\label{fig6}}
\end{figure}
 
Abundances for all the elements are listed in Table 2 and referred to the solar values
adopted from Anders \& Grevesse (1989). We also give \mbox{errors},
 estimated from the dispersion of the elemental abundances, $\Delta_\sigma$,
obtained from the best fits to the various features. We have verified that the
major source for these errors is the inaccuracy in the location of the continuum
caused by the signal-to-noise ratio and the large rotational broadening of
the lines. Errors associated with uncertainties in
effective temperature, $\Delta_{T_{\mathrm{eff}}}$, and gravity, $\Delta{\log
g}$, are also listed in Table 2. The error in the abundance caused by uncertainty
in the determination of the veiling is about 0.05 dex for all the elements.
We have listed in Table 2
the total error, $\Delta_{\rm TOTAL}$, which takes into account all these sources of uncertainty. 

In Fig.\ 3 we show several Fe spectral features where we can see the
best model synthesis in comparison with the synthesis using solar abundances. 
We also analyzed the spectrum of a template
late type star, HD 209100, which was broadened using the rotational profile
determined from the spectrum of \mbox{A0620$-$00}. In the comparison star, the
abundances of all the  elements studied are close to solar except the Li
abundance, which is severely depleted.  
 
Ca and Ni abundances were derived from features where the contribution of
these elements was dominant with little blending from other elements, mostly Fe
lines. Several Ca spectral features are shown in Fig.\ 4; in general these 
features are well reproduced by the synthetic spectra, except for just one,
which was blended with an unidentified line in the solar spectrum, and which  was not used 
in the chemical analysis. Ni lines are displayed
in Fig.\ 5. In particular, the unblended moderately strong Ni~{\scshape i} line at 6643.6
{\AA} is nicely reproduced by the synthetic spectra.    

\begin{figure}[ht!]
  \centering
  \includegraphics[width=11cm,angle=90]{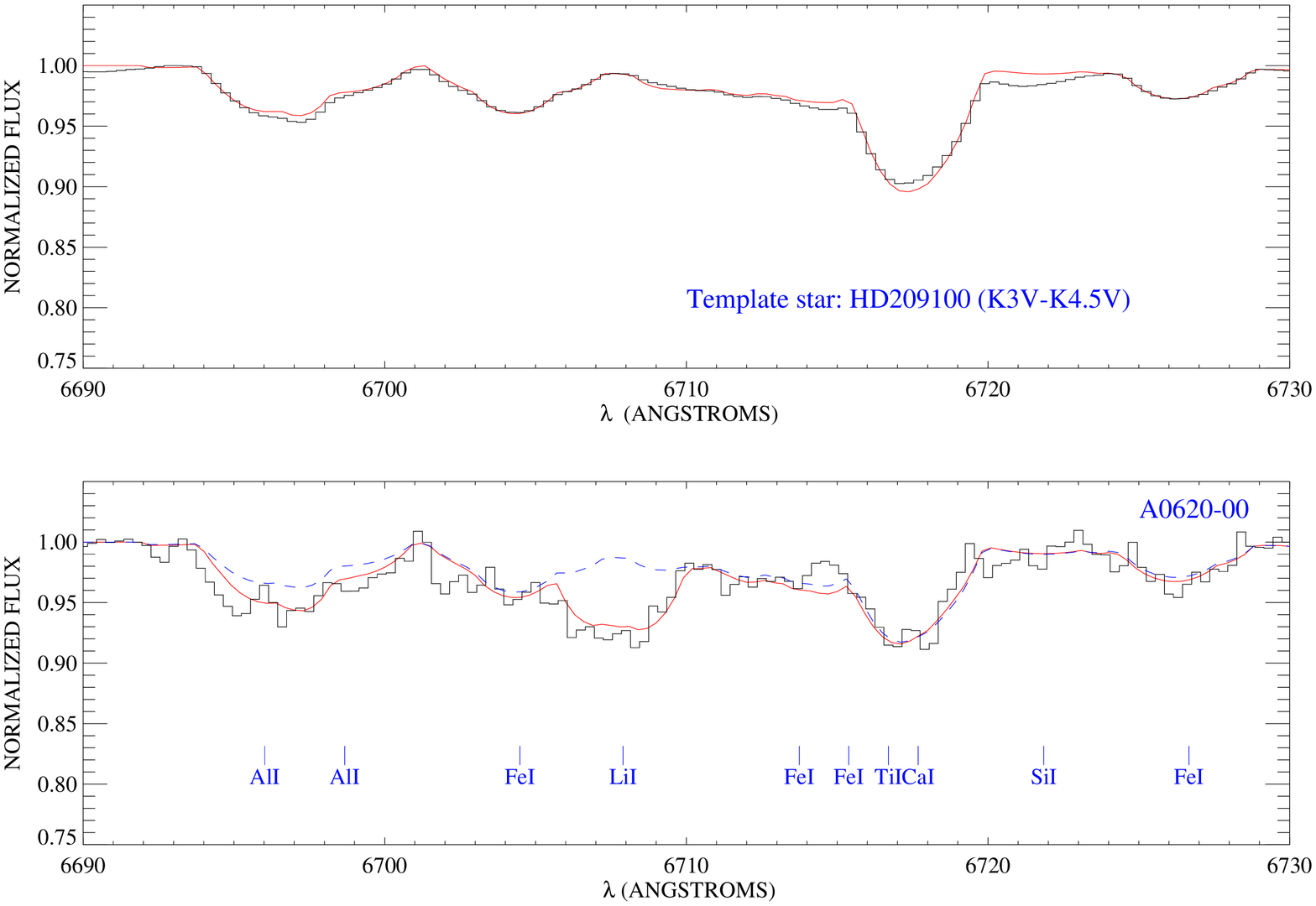}
  \caption{The same as in Fig.\ 3.\label{fig7}}
\end{figure}

The selected Ti lines have been corrected for any possible contamination from
relatively strong telluric lines (EW$\sim 30$ m{\AA} compared to  the
$\sim 150$ m{\AA}  Ti lines). In Fig.\ 6, telluric lines are corrected in
the spectrum of the secondary star in \mbox{A0620$-$00} while in the template
these lines (artificially broadened in the convolution) were not subtracted and
their presence can be clearly seen. 

In Fig.\ 7 we display the only  features of  Al and Li investigated, as well as their
respective spectral synthesis fits. A comparison is also made with the template
star. These are clean weakly blended features where the continuum can be
reliably established. The error in the abundances associated with the 
uncertainty in the continuum location is assumed to be comparable to that of 
single features in other  elements, i.e., 0.1 dex.  Li was clearly enhanced with 
respect the template abundance. The best fit abundance in the template star was $\log \epsilon
(\mathrm{Li}) \le 0.35$ while in the secondary star of \mbox{A0620$-$00} it was $\log
\epsilon (\mathrm{Li}) = 2.31 \pm 0.21$.  

\section{Discussion}

The abundances of {\it heavy} elements in the se\-con\-dary star in
\mbox{A0620$-$00} are slightly higher than solar. We will examine whether these
abundances are anomalous with respect other stars of similar Fe abundance and
what abundance ratios would be expected according to a plausible evolutionary
scenario for the system. 

\subsection{Heavy Elements}

The Fe abundance of the secondary star is slightly higher than solar but
similar to that of many stars in the solar neighborhood. The abundances of
other elements listed in Table 2 have to be understood in the context of the
chemical evolution of the Galaxy. In Fig.\ 8 these element abundances
relative to iron are shown in comparison with the Galactic abundance trends
of these elements in the relevant range of metallicities, taken from Feltzing
\& Gustafsson (1998) and Bodaghee et al. (2003). The error in the element
abundance ratios ($\mathrm{[E/Fe]}$) takes into account how individual
element abundances depend on the various sources of uncertainty. As can be
seen in Table 2, the uncertainties induced by effective temperature and
gravity are considerably diminished when dealing with abundance ratios and
the major source of error in $\mathrm{[E/Fe]}$ is associated with the
dispersion, $\Delta_{\sigma}$, of abundances  obtained  from different features of the same
element. The $\mathrm{[Ca/Fe]}$ ratio of the secondary is
consistent with abundances of stars with similar iron content, while Ni and
Ti appear to be moderately enhanced. Al is clearly over-abundant if we take
into account the low dispersion of Al abundances in these stars. In Table 3
we show the element abundance ratios in the secondary star in A0620$-$00 and
these average values in stars with iron content in the range $-0.06 <
\mathrm{[Fe/H]} < 0.34$, the comparison sample, corresponding to a 1$\sigma$ 
uncertainty in the iron abundance of the companion star. Whereas Ca is
consistent with the average values of the comparison sample, Ni, Ti and
especially Al are 1$\sigma$ more over-abundant than the average values of the
stars in the comparison sample.      

We will discuss these results in the framework of possible evolutionary
scenarios of the \mbox{A0620$-$00} system.  

\subsubsection{Evolutionary scenario}

\begin{figure}[ht!]
\rotate
  \centering
  \includegraphics[width=13cm,angle=0]{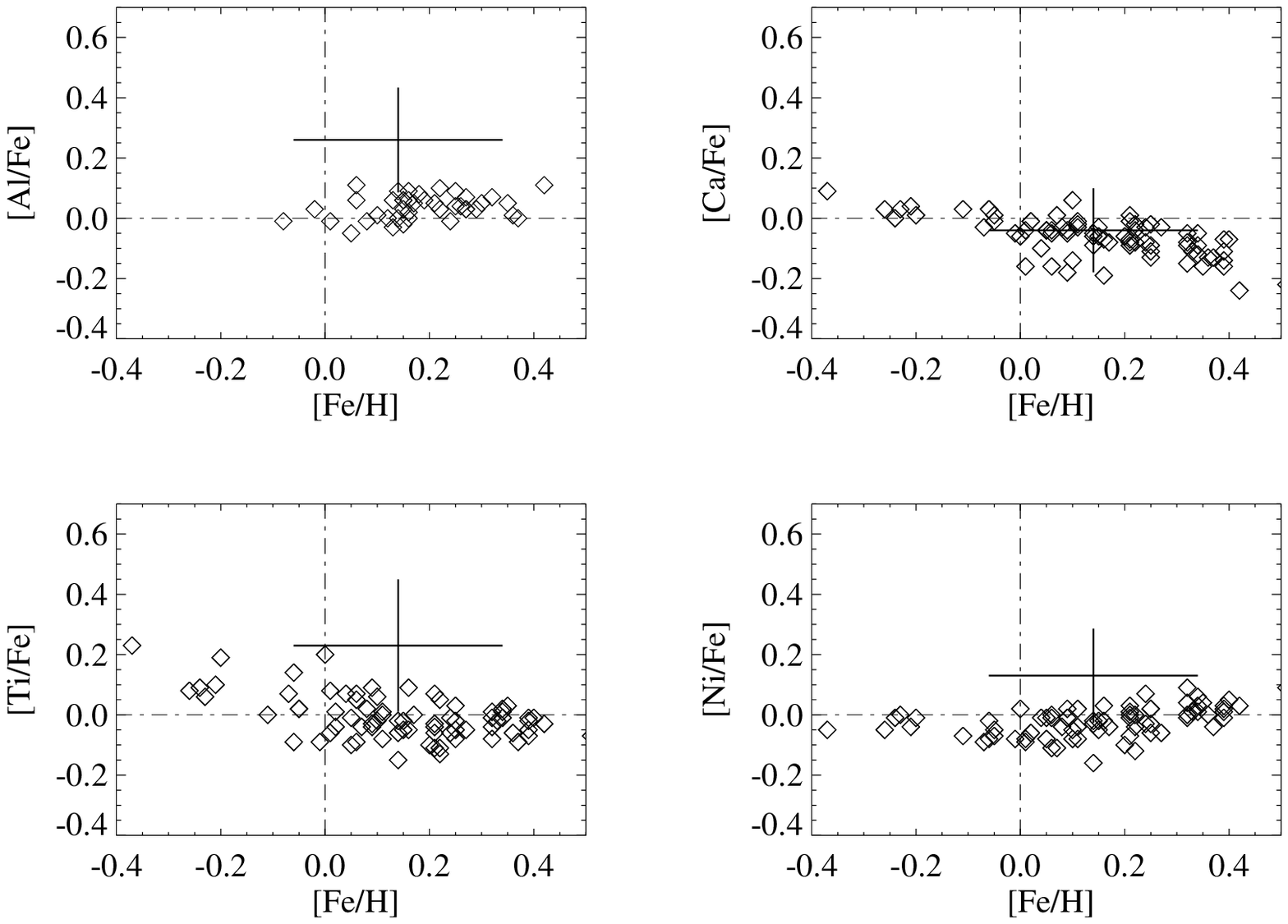}
  \caption{Abundances of the secondary star in \mbox{A0620$-$00} (wide crosses) in comparison with the
  abundances in G and K metal-rich dwarf stars. Trends of Ca, Ni, and Ti were taken from
  Bodaghee et al.\ (2003) while Al from Feltzing \& Gustafsson (1998). The size of the
  cross indicates the error. The dashed-dotted lines indicate solar abundance
  values.\label{fig8}} 
\end{figure} 

\begin{deluxetable}{lccccc}
\tabletypesize{\scriptsize}
\tablecaption{Element abundances ratios in the secondary star in
\mbox{A0620$-$00} and in the comparison sample\label{tbl-3}}   
\tablewidth{0pt}
\tablehead{\colhead{Element} & $\mathrm{[E/Fe]}_{\rm A0620-00}$ &
$\Delta^{\tablenotemark{\star}}_{\rm [E/Fe],A0620-00}$ & $\mathrm{[E/Fe]}_{\rm stars}$ & $\sigma_{\rm
stars}$ & $\Delta_{\sigma,{\rm stars}}$} 
\startdata
Al   &  0.26 & 0.17 &  0.04 & 0.04 & 0.006 \\
Ca   & -0.04 & 0.13 & -0.06 & 0.05 & 0.007 \\
Ti   &  0.23 & 0.21 & -0.02 & 0.07 & 0.009 \\
Ni   &  0.13 & 0.15 & -0.03 & 0.05 & 0.006 \\
\enddata

\tablenotetext{\star}{Errors in the element abundance ratios
($\mathrm{[E/Fe]}$) in the secondary star in A0620-00.}

\tablecomments{$\mathrm{[E/Fe]}_{\rm stars}$ indicate the average values
calculated for stars with iron content in the range -0.06 to 0.34
corresponding to 1$\sigma$ in the $\mathrm{[Fe/H]}$ abundance of the
secondary star in A0620-00. Ca, Ti and Ni for the comparison sample have
been taken from 57 stars in Bodaghee et al.\ (2003) while Al from 36 stars in
Feltzing \& Gustafsson (1998). The uncertainty in the average value of
element abundance ratios in the comparison sample is obtained as
$\Delta_{\sigma,{\rm stars}} =\sigma_{\rm stars}/\sqrt{N}$ where $\sigma_{\rm
stars}$ is the standard deviation of the measurements and N, the number of
stars.}    

\end{deluxetable}

The evolutionary scenario proposed by de Kool et al. (1987) starts with a massive
star ($M_1 \sim 40$ {$M_\odot\; $}) and a companion of roughly $M_2 \sim 1$
{$M_\odot$}. Spiral-in of the secondary during the red supergiant phase leads to the
ejection of the hydrogen-rich envelope and produces a short period helium star
binary. In order to be able to reproduce the high mass ($\sim 10$ {$M_\odot$}) of the compact object in 
\mbox{A0620$-$00} we should consider the so-called {\it Case C} for mass transfer
in which the massive star does not lose its envelope before fi\-ni\-shing its He
core burning (Brown et al., 1999). The mass, $M_{\rm He}$, and radius, $R_{\rm He}$,
of the helium core of the progenitor can be computed using the expressions given
by Portegies Zwart et al. (1997, and references therein): 
$$M_{\rm He}=0.073M_1^{1.42}$$ and $$\log R_{\rm He} = -1.13 + 2.26\log M_{\rm He} -
0.78 (\log M_{\rm He})^2.$$ In that case, the helium core would have $M_{\rm He} \sim
14$ {$M_\odot\; $} and a $R_{\rm He} \sim 2.7$ {$R_\odot$}, and we assume that the mass and
radius of the secondary star are not significantly affected by the spiral-in.
However, Hjellming \& Taam (1991) have found in common envelope phase in
cataclysmic variable systems that a 1.25 {$M_\odot\; $} secondary star 
could accrete $\lesssim 0.1$ {$M_\odot\; $} from the red giant hydrogen-rich envelope
(with C more depleted than N enriched, Marks \& Sarna 1998). Here, we will
consider that the atmosphere of the secondary star is only polluted by
the ejecta in the supernova explosion of the helium core. In this case, what kind
of anomalies can we expect to find at the surface of the secondary?

An important quantity in this discussion is the mass of the compact object. Gelino
et al. (2001) have recently reported an estimated mass  of $11 \pm 1.9$ {$M_\odot$}
for the compact object in \mbox{A0620$-$00};  hence the final remnant mass in
supernova model should be roughly 10 {$M_\odot$}. The mass cuts in supernovae or even
the more energetic hypernovae give rise to compact objects whose masses are not high
enough to explain the high mass black hole in \mbox{A0620$-$00} (Nakamura et al.,
2001; Woosley \& Weaver 1995). Therefore, the final mass of the compact object
had to be generated either from a prompt and direct collapse (collapsar Type I)
or, in a mild explosion with fall-back (collapsar Type II, MacFadyen et al., 2001).
In the latter case, up to $\sim 5$ {$M_\odot\; $} may fall back onto the
collapsed remnant, turning it into a black hole. This makes it quite
difficult that a significant mass fraction of iron could have escaped from
the collapsing ma\-tter in the supernova event because iron is formed in the
inner layers of the star. Israelian et al. (1999) found in Nova Scorpii 94
that while $\alpha$-elements were enhanced Fe was not, in spite of the lower
mass of the compact object ($5.4 \pm 0.3\ M_\odot$, Beer \& Podsiadlowski
2002) as compared with that in \mbox{A0620$-$00}. Thus, the supernova may not
eject any iron, and it therefore seems quite plausible that the slightly
higher than solar Fe abundance of the secondary star in \mbox{A0620$-$00}
reflects its primordial value. However, mixing of ejected material may be
induced by Rayleigh--Taylor instabilities (Kifonidis et al., 2000); hence
heavy elements such as $^{56}$Ni may be conveyed to the outer layers in the
explosion.

On the other hand, the orbital separation between the compact object and the
secondary has been estimated to be $a_{\rm c} = 4.47 \pm 0.27$ {$R_\odot\; $}
(Gelino et al., 2001). We can relate the post-supernova orbital parameters
with the pre-SN parameters assuming an initially circular orbit and
instantaneous spherically symmetric ejection (that is, in a time interval
shorter than the orbital period). These relations are given by van den Heuvel
\& Habets (1984): $a_0=a_{\rm c}\mu_f$ where $a_0$ and $a_{\rm c}$ are the
orbital separation just before the supernova and after tidal circularization
of the post-SN eccentric orbit, respectively, and $\mu_f = (M_{\rm
BH}+M_2)/(M_{\rm He}+M_2)$ where $M_{\rm BH}$ is the compact remnant mass.
Adopting the already mentioned \mbox{values} ($M_{\rm BH} \sim 10\ M_\odot$,
$M_{\rm He} \sim 14\ M_\odot$ and $M_2 \sim 1$ {$M_\odot$}), we find $a_0
\sim 3.3$ {$R_\odot$}. This distance is important since we assume that the
secondary star was outside the He core ($R_{\rm He} \sim 2.7$ {$R_\odot$}) of
the primary before the SN explosion. Therefore, we can estimate the amount of
the ejected material in a spherical explosion that could be captured by the
companion as though coming from a central point. The  explosion of a star as
massive as the primary would have taken place only $\sim 5\times10^6$ yr
after  the formation of the system (Brunish \& Truran 1982). At that time,
the radius of a 1 {$M_\odot\; $} secondary star would be $\sim 1.3$
{$R_\odot\; $} (D'Antona \& Mazzitelli 1994). If we consider a spherically
symmetric supernova explosion, taking into account the fraction of solid the
angle subtended by the companion and assuming a capture efficiency, $f_{\rm
capture}$, of 1 (i.e., all the matter ejected within that solid angle is
captured), the amount of mass deposited on the secondary would have been
$$m_{\rm add}=\Delta M (\pi R_2^2/4 \pi a_0^2)f_{\rm capture} \sim 0.15\;
M_\odot,$$ where $\Delta M = M_{\rm He}-M_{\rm BH} \sim 4$ {$M_\odot\; $} is
the total ejected mass. This amount of  ejected mass is sufficient to explain
the radial velocity of the system which is a lower limit to its
\emph{runaway} velocity (Nelemans et al., 1999).     

\begin{deluxetable}{lcccccccccccccccccccc}
\centering
\tabletypesize{\scriptsize}
\rotate
\tablecolumns{19}
\tablecaption{Expected element abundances of the secondary star in
\mbox{A0620$-$00}\label{tbl-3}}  
\tablewidth{0pc}
\tablehead{Element & ${\rm [E/H]}$\tablenotemark{\ddagger} & ${\rm [E/H]}$\tablenotemark{\dagger} & \multicolumn{16}{c}{${\rm [E/H]}$\tablenotemark{\star}}}
\startdata
\noalign{\smallskip}
$\varepsilon{\rm (}10^{51}{\rm erg)}$ & & & \multicolumn{8}{c}{$1$} & \multicolumn{8}{c}{$30$} \\
\noalign{\smallskip}
\hline
\noalign{\smallskip}
$M_{\rm cut}$({$M_\odot$}) & & & 1.96 & 5 & 7 & 10 & 11 & 12 & 12.5 & 12.9 &
2.03 & 5 & 7 & 10 & 11 & 12 & 12.5 & 12.9\\ 
\noalign{\smallskip}
\hline
\noalign{\smallskip}
$M_2$ & & & \multicolumn{16}{c}{1 $M_\odot\; $} \\ 
\noalign{\smallskip}
\hline
\noalign{\smallskip}
Al & 0.40 & 0.21 & 1.50 &  1.52 &  1.50 &  1.40 &  1.12 &  0.68 &  0.37 &  0.22 &  1.36 &  1.47 &  1.51 &  1.40 &  1.12 &  0.70 &  0.38 &  0.22 \\
Ca & 0.10 & 0.09 & 0.88 &  0.12 &  0.12 &  0.13 &  0.12 &  0.11 &  0.11 &  0.10 &  1.26 &  0.14 &  0.12 &  0.12 &  0.11 &  0.10 &  0.10 &  0.10 \\
Ti & 0.37 & 0.13 & 0.82 &  0.31 &  0.31 &  0.29 &  0.23 &  0.18 &  0.16 &  0.15 &  1.47 &  0.31 &  0.31 &  0.30 &  0.23 &  0.18 &  0.16 &  0.15 \\
Fe & 0.14 & 0.14 & 0.94 &  0.17 &  0.17 &  0.17 &  0.17 &  0.16 &  0.16 &  0.15 &  1.26 &  0.16 &  0.16 &  0.17 &  0.16 &  0.15 &  0.15 &  0.15 \\
Ni & 0.27 & 0.12 & 1.12 &  0.49 &  0.48 &  0.44 &  0.33 &  0.23 &  0.18 &  0.15 &  1.47 &  0.54 &  0.52 &  0.52 &  0.41 &  0.31 &  0.27 &  0.23 \\
\noalign{\smallskip}
\hline
\noalign{\smallskip}
 O & \nodata & \nodata & 1.16 &  1.18 &  1.17 &  1.11 &  0.90 &  0.61 &  0.39 &  0.15 &  1.12 &  1.20 &  1.17 &  1.11 &  0.89 &  0.61 &  0.39 &  0.15 \\
Mg & \nodata & \nodata & 1.20 &  1.18 &  1.16 &  1.07 &  0.83 &  0.50 &  0.30 &  0.15 &  1.13 &  1.24 &  1.21 &  1.10 &  0.87 &  0.57 &  0.39 &  0.25 \\
Si & \nodata & \nodata & 1.05 &  0.41 &  0.40 &  0.36 &  0.27 &  0.19 &  0.16 &  0.15 &  1.35 &  1.04 &  0.54 &  0.37 &  0.27 &  0.19 &  0.17 &  0.15 \\
 S & \nodata & \nodata & 0.93 &  0.19 &  0.19 &  0.19 &  0.18 &  0.16 &  0.16 &  0.15 &  1.32 &  0.91 &  0.22 &  0.19 &  0.17 &  0.16 &  0.15 &  0.15 \\
 C & \nodata & \nodata & 0.82 &  0.87 &  0.86 &  0.84 &  0.68 &  0.50 &  0.37 &  0.16 &  0.69 &  0.78 &  0.85 &  0.84 &  0.68 &  0.50 &  0.37 &  0.17 \\
\noalign{\smallskip}
\hline
\noalign{\smallskip}
$M_2$ & & & \multicolumn{16}{c}{0.8 $M_\odot\; $}\\ 
\noalign{\smallskip}
\hline
\noalign{\smallskip}
Al & 0.40 & 0.21 & 1.43 &  1.45 &  1.43 &  1.33 &  1.05 &  0.63 &  0.35 &  0.22 &  1.29 &  1.40 &  1.43 &  1.33 &  1.06 &  0.65 &  0.36 &  0.22 \\
Ca & 0.10 & 0.09 & 0.82 &  0.12 &  0.12 &  0.12 &  0.11 &  0.11 &  0.10 &  0.10 &  1.19 &  0.13 &  0.11 &  0.12 &  0.11 &  0.10 &  0.10 &  0.10 \\
Ti & 0.37 & 0.13 & 0.76 &  0.29 &  0.29 &  0.27 &  0.22 &  0.17 &  0.16 &  0.14 &  1.40 &  0.29 &  0.29 &  0.27 &  0.22 &  0.17 &  0.16 &  0.15 \\
Fe & 0.14 & 0.14 & 0.88 &  0.16 &  0.16 &  0.17 &  0.16 &  0.16 &  0.15 &  0.15 &  1.19 &  0.16 &  0.16 &  0.16 &  0.15 &  0.15 &  0.15 &  0.15 \\
Ni & 0.27 & 0.12 & 1.05 &  0.45 &  0.44 &  0.41 &  0.30 &  0.21 &  0.17 &  0.14 &  1.40 &  0.50 &  0.48 &  0.48 &  0.38 &  0.29 &  0.25 &  0.22 \\
\noalign{\smallskip}
\hline
\noalign{\smallskip}
 O & \nodata & \nodata & 1.10 &  1.11 &  1.10 &  1.05 &  0.83 &  0.56 &  0.36 &  0.14 &  1.05 &  1.14 &  1.10 &  1.04 &  0.83 &  0.56 &  0.36 &  0.14 \\
Mg & \nodata & \nodata & 1.13 &  1.11 &  1.09 &  1.01 &  0.77 &  0.46 &  0.28 &  0.15 &  1.06 &  1.17 &  1.14 &  1.04 &  0.81 &  0.53 &  0.36 &  0.24 \\
Si & \nodata & \nodata & 0.98 &  0.38 &  0.37 &  0.34 &  0.25 &  0.18 &  0.16 &  0.15 &  1.28 &  0.97 &  0.50 &  0.34 &  0.26 &  0.19 &  0.16 &  0.15 \\
 S & \nodata & \nodata & 0.87 &  0.18 &  0.18 &  0.19 &  0.17 &  0.16 &  0.15 &  0.15 &  1.25 &  0.85 &  0.21 &  0.18 &  0.17 &  0.16 &  0.15 &  0.15 \\
 C & \nodata & \nodata & 0.76 &  0.81 &  0.80 &  0.79 &  0.62 &  0.46 &  0.34 &  0.16 &  0.64 &  0.72 &  0.79 &  0.78 &  0.62 &  0.46 &  0.34 &  0.16 \\
\enddata
\tablenotetext{\star}{Secondary star abundances relative to solar.}  

\tablenotetext{\ddagger}{This column shows observed abundances of the secondary star in
A0620$-$00.}

\tablenotetext{\dagger}{This column shows the average abundances in stars of
the comparison sample (see also Table 3).}

\tablecomments{Expected abundances in the secondary atmosphere contaminated
with nuclesynthetic products of a 40 {$M_\odot\; $} spherically symmetric
core-collapse explosion 
model ($M_{\rm He} \sim 14\ M_\odot$) for two different explosion energies.
$M_{\rm cut}$ is the mass cut assumed for each model. In every model, the
final remnant mass, $M_{\rm BH}$, is fixed at 10 {$M_\odot$}, except those
models with $M_{\rm cut}=11$, 12, 12.5, and $12.9$ {$M_\odot$}. Mixing factors between the fall-back matter and the ejecta have
been adopted equal to 1. In every model, the amount of fall-back, $M_{\rm
fallback}$, is the difference between $M_{\rm BH}$ and $M_{\rm cut}$. The
capture efficiency, $f_{\rm capture}$, is fixed at 1. Elements which have not
been analyzed (i.e. O, Mg, S, Si and C) have been scaled up with ${\rm 
[Fe/H]}_{\dagger,i}=0.14$.}   

\end{deluxetable}

\begin{deluxetable}{lcccccccccccccccccccc}
\centering
\tabletypesize{\scriptsize}
\rotate
\tablecolumns{19}
\tablecaption{Expected element abundances of the secondary star in \mbox{A0620$-$00}\label{tbl-3}}  
\tablewidth{0pc}
\tablehead{Element & ${\rm [E/H]}$\tablenotemark{\ddagger} & ${\rm [E/H]}$\tablenotemark{\dagger} & \multicolumn{16}{c}{${\rm [E/H]}$\tablenotemark{\star}}}
\startdata
\noalign{\smallskip}
$\varepsilon{\rm (}10^{51}{\rm erg)}$ & & & \multicolumn{8}{c}{$1$} & \multicolumn{8}{c}{$30$} \\
\noalign{\smallskip}
\hline
\noalign{\smallskip}
$M_{\rm cut}$({$M_\odot$}) & & & 1.96 & 5 & 7 & 10 & 11 & 12 & 12.5 & 12.9 &
2.03 & 5 & 7 & 10 & 11 & 12 & 12.5 & 12.9\\ 
\noalign{\smallskip}
\hline
\noalign{\smallskip}
$f_{\rm capture}$ & & & \multicolumn{16}{c}{0.5} \\ 
\noalign{\smallskip}
\hline
\noalign{\smallskip}
Al & 0.40 & 0.21 & 1.15 &  1.18 &  1.15 &  1.06 &  0.81 &  0.47 &  0.28 &  0.21 &  1.02 &  1.13 &  1.16 &  1.06 &  0.81 &  0.48 &  0.29 &  0.22 \\
Ca & 0.10 & 0.09 & 0.59 &  0.10 &  0.10 &  0.11 &  0.10 &  0.10 &  0.10 &  0.10 &  0.92 &  0.11 &  0.10 &  0.10 &  0.10 &  0.10 &  0.09 &  0.09 \\
Ti & 0.37 & 0.13 & 0.55 &  0.22 &  0.22 &  0.21 &  0.18 &  0.15 &  0.14 &  0.14 &  1.12 &  0.22 &  0.22 &  0.21 &  0.18 &  0.15 &  0.14 &  0.14 \\
Fe & 0.14 & 0.14 & 0.65 &  0.15 &  0.15 &  0.15 &  0.15 &  0.15 &  0.15 &  0.15 &  0.92 &  0.15 &  0.15 &  0.15 &  0.15 &  0.14 &  0.14 &  0.14 \\
Ni & 0.27 & 0.12 & 0.80 &  0.32 &  0.31 &  0.29 &  0.23 &  0.17 &  0.15 &  0.14 &  1.12 &  0.35 &  0.34 &  0.34 &  0.27 &  0.22 &  0.19 &  0.18 \\
\noalign{\smallskip}
\hline
\noalign{\smallskip}
 O & \nodata & \nodata & 0.84 &  0.86 &  0.84 &  0.80 &  0.61 &  0.40 &  0.26 &  0.14 &  0.80 &  0.88 &  0.84 &  0.79 &  0.61 &  0.40 &  0.26 &  0.14 \\
Mg & \nodata & \nodata & 0.87 &  0.86 &  0.84 &  0.76 &  0.56 &  0.33 &  0.21 &  0.15 &  0.81 &  0.90 &  0.88 &  0.79 &  0.59 &  0.38 &  0.26 &  0.19 \\
Si & \nodata & \nodata & 0.74 &  0.28 &  0.27 &  0.25 &  0.20 &  0.16 &  0.15 &  0.15 &  1.00 &  0.73 &  0.36 &  0.25 &  0.20 &  0.16 &  0.15 &  0.15 \\
 S & \nodata & \nodata & 0.64 &  0.16 &  0.16 &  0.16 &  0.16 &  0.15 &  0.15 &  0.15 &  0.98 &  0.63 &  0.18 &  0.16 &  0.15 &  0.15 &  0.15 &  0.14 \\
 C & \nodata & \nodata & 0.55 &  0.59 &  0.59 &  0.57 &  0.45 &  0.33 &  0.25 &  0.15 &  0.46 &  0.52 &  0.58 &  0.57 &  0.45 &  0.33 &  0.25 &  0.15 \\
\noalign{\smallskip}
\hline
\noalign{\smallskip}
$f_{\rm capture}$ & & & \multicolumn{16}{c}{0.1} \\ 
\noalign{\smallskip}
\hline
\noalign{\smallskip}
Al & 0.40 & 0.21 & 0.62 &  0.63 &  0.62 &  0.55 &  0.41 &  0.28 &  0.22 &  0.21 &  0.53 &  0.60 &  0.62 &  0.56 &  0.41 &  0.28 &  0.23 &  0.21 \\
Ca & 0.10 & 0.09 & 0.25 &  0.09 &  0.09 &  0.09 &  0.09 &  0.09 &  0.09 &  0.09 &  0.42 &  0.09 &  0.09 &  0.09 &  0.09 &  0.09 &  0.09 &  0.09 \\
Ti & 0.37 & 0.13 & 0.25 &  0.15 &  0.15 &  0.15 &  0.14 &  0.14 &  0.13 &  0.13 &  0.57 &  0.15 &  0.15 &  0.15 &  0.14 &  0.14 &  0.13 &  0.13 \\
Fe & 0.14 & 0.14 & 0.30 &  0.14 &  0.14 &  0.14 &  0.14 &  0.14 &  0.14 &  0.14 &  0.44 &  0.14 &  0.14 &  0.14 &  0.14 &  0.14 &  0.14 &  0.14 \\
Ni & 0.27 & 0.12 & 0.37 &  0.18 &  0.17 &  0.17 &  0.15 &  0.14 &  0.14 &  0.13 &  0.57 &  0.19 &  0.18 &  0.18 &  0.16 &  0.15 &  0.15 &  0.14 \\
\noalign{\smallskip}
\hline
\noalign{\smallskip}
 O & \nodata & \nodata & 0.40 &  0.40 &  0.40 &  0.37 &  0.28 &  0.21 &  0.17 &  0.14 &  0.38 &  0.42 &  0.40 &  0.37 &  0.28 &  0.21 &  0.17 &  0.14 \\
Mg & \nodata & \nodata & 0.41 &  0.41 &  0.39 &  0.35 &  0.26 &  0.18 &  0.16 &  0.14 &  0.38 &  0.43 &  0.42 &  0.37 &  0.28 &  0.20 &  0.17 &  0.15 \\
Si & \nodata & \nodata & 0.34 &  0.17 &  0.17 &  0.16 &  0.15 &  0.14 &  0.14 &  0.14 &  0.50 &  0.34 &  0.19 &  0.16 &  0.15 &  0.14 &  0.14 &  0.14 \\
 S & \nodata & \nodata & 0.30 &  0.14 &  0.14 &  0.14 &  0.14 &  0.14 &  0.14 &  0.14 &  0.48 &  0.29 &  0.15 &  0.14 &  0.14 &  0.14 &  0.14 &  0.14 \\
 C & \nodata & \nodata & 0.26 &  0.28 &  0.27 &  0.27 &  0.22 &  0.18 &  0.16 &  0.14 &  0.22 &  0.25 &  0.27 &  0.27 &  0.22 &  0.18 &  0.17 &  0.14 \\
\enddata
\tablenotetext{\star}{Secondary star abundances relative to solar.}  

\tablenotetext{\ddagger}{This column shows observed abundances of the secondary star in
A0620$-$00.}

\tablenotetext{\dagger}{This column shows the average abundances in stars of
the comparison sample (see also Table 3).}

\tablecomments{The same as in Table 4 but the secondary mass is fixed at 0.8
$M_\odot$. However, the capture efficiency, $f_{\rm capture}$, is lower
than 1. Its values are 0.5 which means that only half of the matter ejected 
within the solid angle subtended by the companion is captured whereas 0.1
means that only 10 per cent is captured.}  

\end{deluxetable}

In Tables 4 and 5 we present the expected abundances in the atmosphere 
of the secondary star after contamination from the progenitor of the 
compact object according to the above mentioned assumptions. In Table 4 we
have considered two plausible masses (1 {$M_\odot\; $} and 0.8 {$M_\odot$})
for the secondary star at the time of the explosion of the primary 
and a capture efficiency factor of 1, which means that all the matter ejected
within the solid angle subtended by the companion is captured. In Table 5 we
have fixed the secondary mass at 0.8 {$M_\odot\; $} and we have considered
two other efficiency factors of 0.5 and 0.1, which means that 50 and 10 per
cent of the ejected mass within that solid angle is captured. We have used a
40 {$M_\odot\; $} spherically symmetric core-collapse explosion model ($M_{\rm He} \sim 14\
M_\odot$) for two different explosion energies from Umeda \& Nomoto (2002 and
2003, private communication). That energy is deposited instantaneously in the
central region of the progenitor core to generate a strong shock wave. The
subsequent propagation of the shock wave is followed through a hydrodynamic
code (Umeda \& Nomoto, 2002, and references therein). In our simple model, we
have assumed different mass cuts, fall-back masses and a mixing
factor\footnote{Other detailed explosion models have been considered in the 
analysis of over-abundances of the secondary star in Nova Sco 94 by
Podsiadlowski {\it et al.}\ (2002).} that take into account the amount of
fall-back matter mixed with the ejecta. We have adopted a mixing factor of 1
(i.e., all the fall-back material is well mixed with the ejecta). The amount
of fall-back, $M_{\rm fallback}$, is the difference between the final remnant
mass, $M_{\rm BH}$, and the initial remnant mass, $M_{\rm cut}$. In models
with $M_{\rm cut} \le 10\ M_\odot$, the final remnant mass is equal to 10
{$M_\odot\; $} and hence the total ejected mass is 4 $M_\odot$. Nevertheless,
models with $M_{\rm cut} > 10\ M_\odot$ assume $M_{\rm BH} = M_{\rm cut}$ and
do not include either fall-back or mixing. In such models, the total ejected 
mass, $\Delta M$, is the diference between the final remnant mass and the
helium core mass.   

At the time of the explosion, the secondary was close enough to the  
He core ($a_0 \sim 3.3$ {$R_\odot$}) for the amount of matter that could have been
accreted from the H-He enriched envelope after explosion to be negligible, so 
we do not expect a significant change in the He/H ratio in the secondary atmosphere.
The hydrogen mass in the He core was $\sim 10^{-16}$ {$M_\odot\; $}; therefore, we can
estimate the number density of an element E in the secondary atmosphere as:  

 $$\left[\frac{N({\rm E})}{N({\rm H})}\right]_{\star,f} = \left[\frac{N({\rm
 E})_\star+N({\rm E})_{\rm add}}{N({\rm H})_\star}\right]= 
$$$$\left[\frac{X({\rm E})_\star}{X({\rm H})_\star}10^{{\rm
[E/H]}_{\dagger,i}}+\frac{X({\rm E})_{\rm add}}{X({\rm H})_\star}\frac{m_{\rm
add}}{m_{\rm conv}}\right]\frac{m_{\rm H}}{m_{\rm E}},$$ where $N({\rm E})_\star$ and $N({\rm H})_\star$ are
the number density of elements E and H in the convective zone of the star,
$N({\rm E})_{\rm add}$ the number density of E in the captured matter, $m_{\rm
add}$, by the companion, $X({\rm E})_\star$ and $X({\rm H})_\star$ the mass
fractions of E and H in the convective zone, $X({\rm E})_{\rm add}$ is the mass
fraction of E in the captured matter, $m_{\rm conv}$ is the mass of the
convective zone, and $m_{\rm H}$ and $m_{\rm E}$ the atomic masses of H and E,
respectively. The mass of the convective zone was fixed at $m_{\rm conv} =
0.652$ and $0.667$ {$M_\odot\; $} from the evolutionary tracks for  a 1 and
0.8 {$M_\odot\; $} secondary stars, respectively, with an age of $\sim
5\times10^6$ yr (D'Antona \& Mazzitelli 1994).   

The original element mass fractions in the secondary star were assumed to be
solar from Anders \& Grevesse (1989) and we have also considered an
explosion model from a solar metallicity progenitor. However, in
order to take into account that the initial iron content of the star was
probably higher than solar (${\rm [Fe/H]}_{\dagger,i}=0.14$) we scaled up the
expected abundances with ${\rm[E/H]}_{\dagger,i}$ which is the average element
abundance in the comparison sample using data shown in Fig.\ 8 (see Table 3).
Expected abundances shown in Table 4 and 5 were calculated using the
following formula:      

$${\rm [E/H]}_{\star,f} = \log \left[\frac{N({\rm E})}{N({\rm
H})}\right]_{\star,f} - \log \left[\frac{N({\rm E})}{N({\rm
H})}\right]_{\odot}$$  
 
\begin{figure}[ht!]
  \centering
  \includegraphics[width=13cm,angle=0]{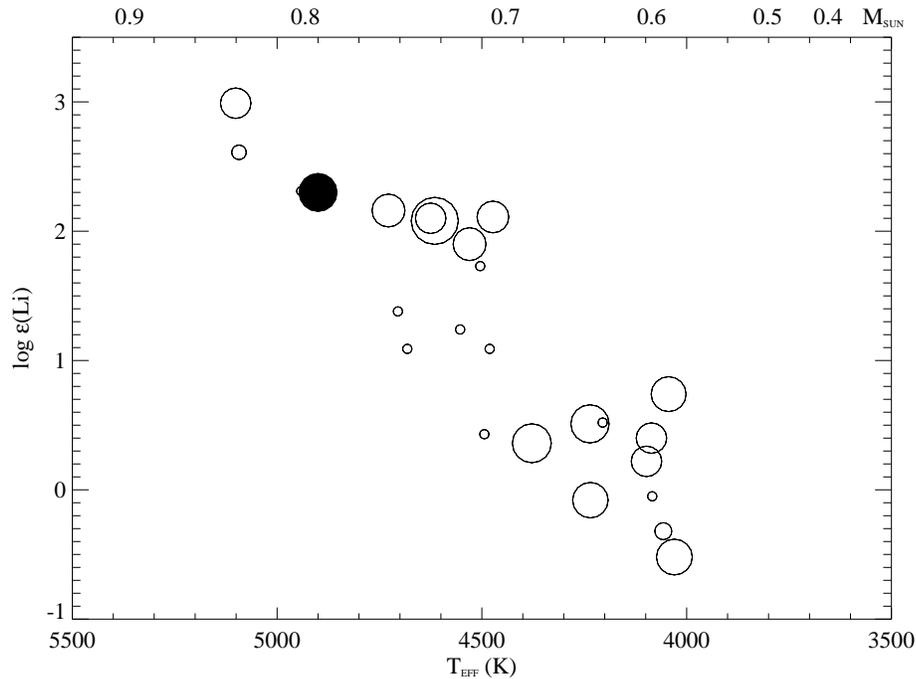}
  \caption{Li abundance of the secondary star in \mbox{A0620$-$00} (filled circle)
  in comparison with the abundances of different rotating Pleiades dwarf stars
  versus effective temperature from Garc{\'\i }a L\'opez et al. (1994). The sizes of
  the circles are related to $v \sin i$. Stellar masses have been assigned
  following the evolu\-tio\-na\-ry tracks of D'Antona \& Mazzitelli (1994) for
  $10^8$ yr.\label{fig9}}  
\end{figure} 

It is clearly seen that only models with mass cuts as high as 11
{$M_\odot\; $} can fit the abundances of the secondary star within the 
error bars. In models with lower mass cuts, the amount of Al in the ejecta is
too high in comparison with the other heavy elements analyzed. Even under
strong assump\-tions such as large amounts of fall-back matter, $M_{\rm
fallback}$, and/or mixing efficiency fixed at 1, it is not possible to get
good fits to the observed abundance values for mass cuts below 11--12
{$M_\odot$} depending on the capture efficiency assumed. This is independent
of the explosion energy, $\varepsilon$, since this parameter has little
influence on the Al yield at any mass cut. It does not depend on the mass of
the secondary. In summary, the moderate over-abundances of Ti, Ni, and
especially Al could be explained if there were an explosion event of a star
of initial mass $\sim 40$ {$M_\odot\; $} with a helium core of $\sim 14$
{$M_\odot\; $} that led to the formation of a black hole with a mass of
approximately 11--12.5 {$M_\odot\; $}. As in the case of Nova Sco 94
(Israelian et al., 1999), we find signatures in the secondary of
\mbox{A0620$-$00} that suggest the formation  of the black hole as a
consequence of an explosive event. 

\begin{figure}[ht!]
  \centering
  \includegraphics[width=13cm,angle=0]{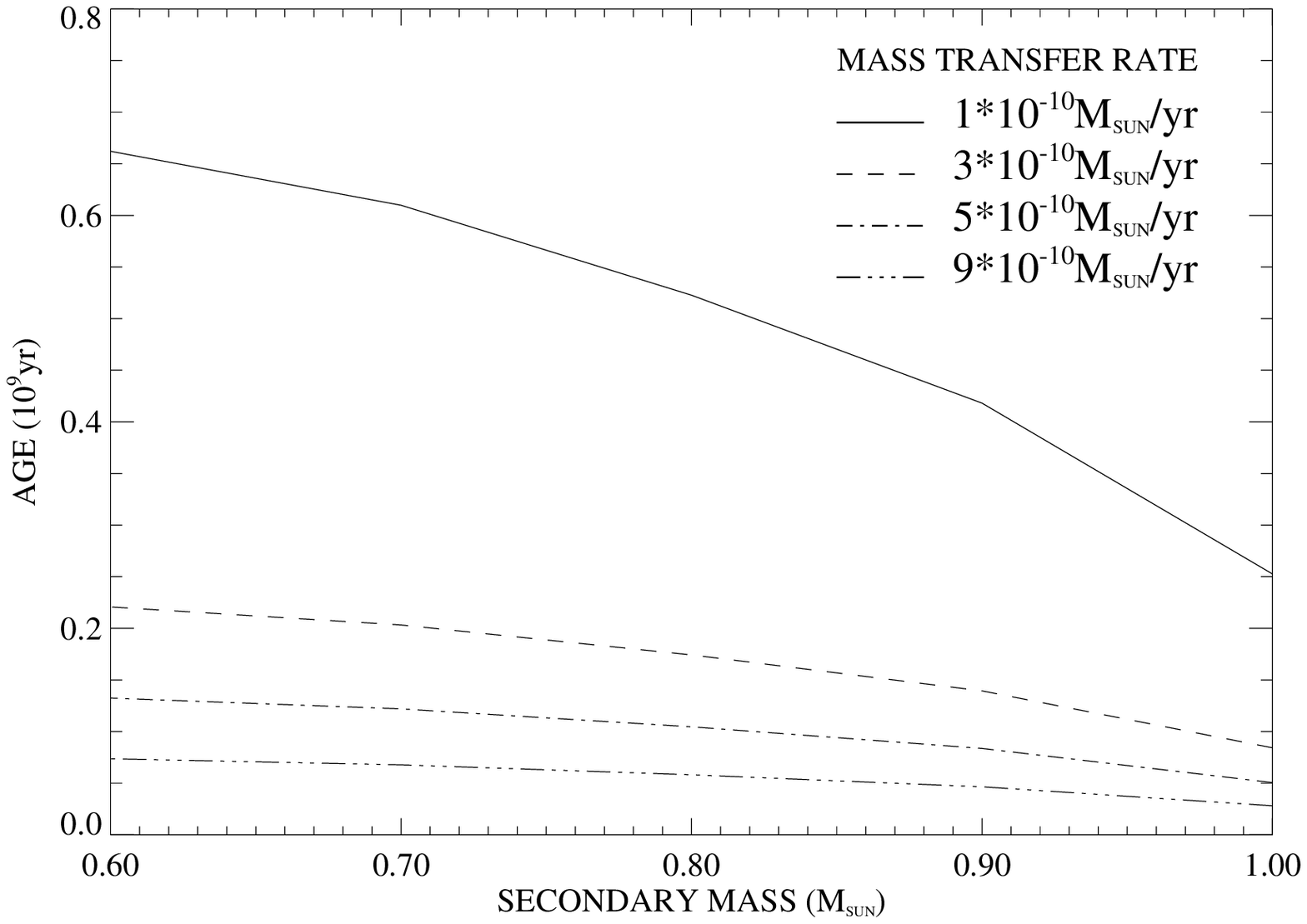}
  \caption{Upper limit to the age of the A0620$-$00 system according to the Li
  abundance of the secondary star versus stellar masses of the secondary star,
  assuming different mass transfer rates. Stellar masses have been assigned
  according to the evolu\-tio\-na\-ry tracks of D'Antona \& Mazzitelli (1994) for
  $10^8$ yr.\label{fig9}}  
\end{figure} 

\begin{figure}[ht!]
  \centering
  \includegraphics[width=11cm,angle=90]{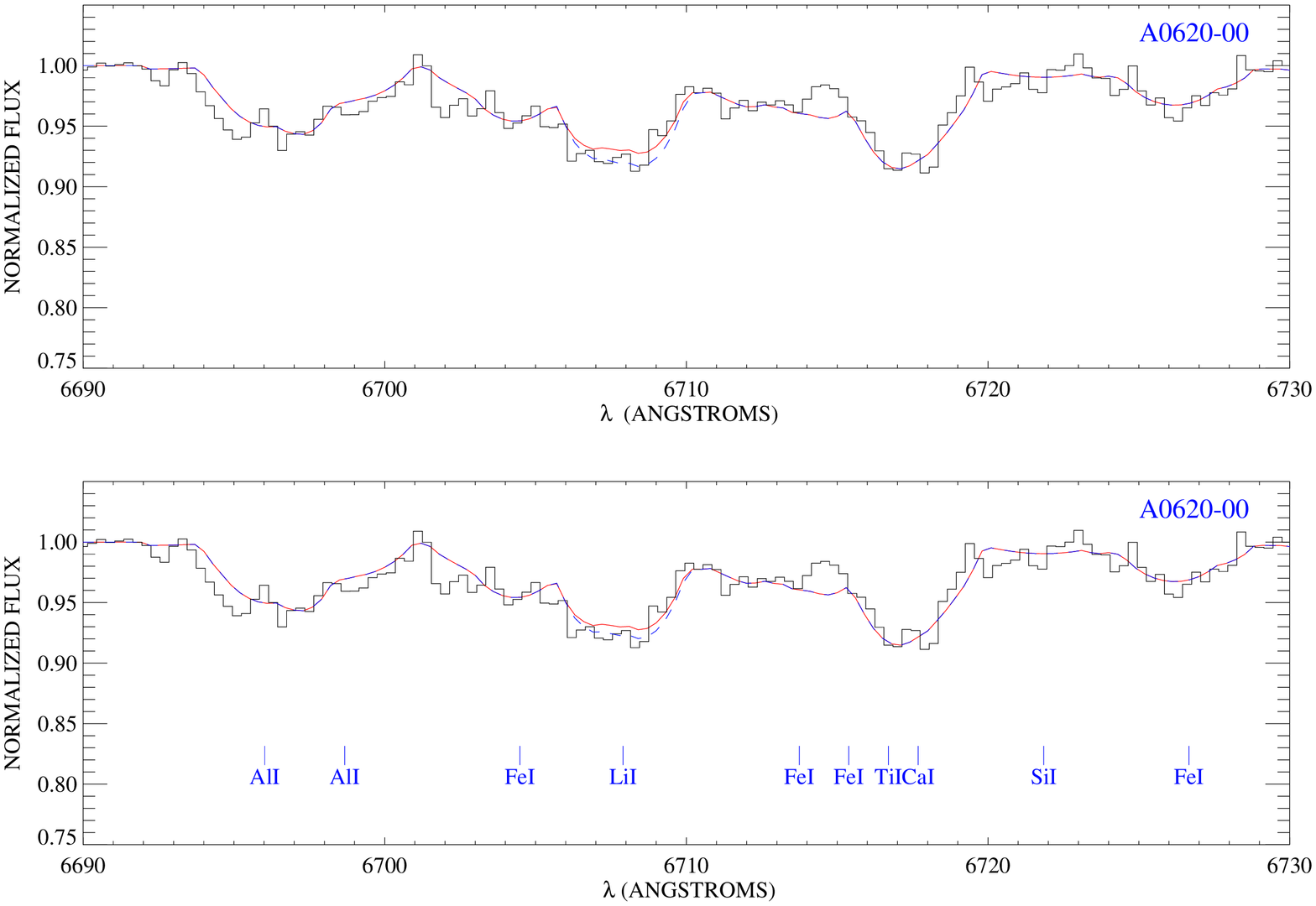}
  \caption{Best synthetic spectral fits to the UVES spectrum of the secondary
    star in the \mbox{A0620-00} system considering only $^7$Li in the spectral line
    Li~{\scshape i}6708. Superimposed to them were displayed different spectral synthesis,
    fixed Li abundance, for $^7{\rm Li}/^6{\rm Li}=2$ (dashed line in top panel) and for 
    $^7{\rm Li}/^6{\rm Li}=5$ (dashed line in bottom panel).\label{fig10}}
\end{figure}

However, as we have already mentioned in \S 1, many uncertainties
still remain in supernova explosion models and in the evolution of LMXBs. For
instance, a capture efficiency as high as 1 may not be adequate since
simulations of type Ia SNe suggest that the supernova blast wave may induce
mass-loss in the secondary instead of matter accretion (Marietta et al.,
2000). On the other hand, the assumption of spherical symmetry is not
generally expected for hypernova or collapsar models, in particular those
associated with gamma-ray burst (MacFadyen \& Woosley, 1999). Maeda et al.
(2002) have studied nucleosynthesis in aspherical explosions and 
found that the chemical composition of the ejecta is strongly dependent on
direction. In particular, Fe is mainly ejected in the polar direction,
whereas O and other alpha-elements (e.g. Si, S, Mg) are preferentially ejected
near the equatorial plane. Therefore, it would be interesting to ana\-ly\-ze
other elements such as O, Mg, and C, which are enhanced in many of the model
computations in Table 4 and 5, even at high mass cuts. We note that Orosz et
al. (2001) have also found overabundances (by a factor 2--10 with respect to
solar) of some of these elements in the secondary star of the LMXB system
J1819.3$-$2525 (V4641 Sgr).   

\subsection{Li abundance}
 
The abundance of lithium in the secondary star in A0620$-$00 is substantially
higher than in field main sequence stars of the same mass ($\sim$ 0.7
M$_{\odot}$) but similar to that of Pleiades stars of comparable mass and
rotational velocity (see Fig.\ 9). Convective mixing during pre-main sequence
and main sequence evolution is expected to produce significant lithium depletion
in the atmospheres of such stars, thus a possible explanation for the Li
over-abundance in the secondary is that the A0620$-$00 system is as young as the
Pleiades cluster(i.e., $\sim$ 7--15 $\times 10^7$ yr). This appears to be in conflict
with the evolutionary scenario proposed by de Kool et al. (1987), where an age of a
few times 10$^9$ yr is assumed. However, Naylor \& Podsiadlowski (1993) have
argued, based on the galactic distribution of LMXBs, that these systems are
associated with the Galactic disk, and probably have an age of only $10^7$ to
$10^8$ yr. If the lifetime were this short, the mass loss from the companion
would not have  been relevant since the present mass transfer rate is $\dot{M}_{2}
\sim 10^{-10}$ {$M_\odot\; $} yr$^{-1}$ (McClintock et al., 1995) although it may not
necessarily be stable in the course of the whole binary lifetime ($\dot{M}_{2}
\sim 10^{-9}$--$10^{-10}$ {$M_\odot\; $} yr$^{-1}$, Ergma \& Fedorova 1998). 

It is possible to constrain the age of the system
using  the  current Li abundance  and  plausible mass transfer rates. Let us
assume in what follows that there is no mechanism able to enrich the atmosphere 
of the secondary star with freshly synthesized Li nuclei.
 If the initial mass of the secondary
star were smaller than 1  $M_{\odot}$ Li could survive only in the outer layers,
i.e., above the bottom of the convective zone. In a solar type  star, 
the mass of the convective zone is 
0.03 $M_{\odot}$, and, given the range of plausible mass transfer rates for the
secondary, such a mass would be transferred to the compact object in less than
0.3 Gyr, leaving the atmosphere of the  secondary completely free of Li nuclei,
which we see cannot be the case.
In Fig.\ 10 we show how this simple argument constrains the age of the system
for different mass transfer rates and initial secondary masses.
In summary, the more massive the secondary star and the higher the mass transfer
rate are, the  stringent the  upper limit to the age of the system that can be
imposed. The curves in 
Fig.\ 10, in fact, give  conservative upper limits to the age, in particular for the lower 
mass range, since we have not taken into account that some Li depletion  caused by convective mixing 
may  also have occurred. We note that we have not considered values for the initial mass
of the secondary above 1 $M_{\odot}$ because under the mass transfer rates considered
such stars cannot lead to an object of the present mass and preserve the required
amount of Li. 

High Li abundances have been noticed by Mart{\'\i }n et al. (1994) in other
LMXBs such as \mbox{Cen X-4} and V404 Cyg. These authors suggested as an alternative
explanation of  youth  the existence of a  production mechanism  of  lithium. Either 
 early in the evolution of the system during  the supernova
explosion of the primary progenitor; or a continuing process such as $\alpha$--$\alpha$
reactions during the repeated strong outbursts that cha\-rac\-te\-ri\-ze
transient X-ray binaries. If, indeed,  fresh Li is synthesized and trapped in the
atmosphere, the above arguments to constrain the age cannot hold. It is thus, very
important to find a way of disentangling the origin of the high Li abundances
observed.
The spallation mechanism would produce considerable amounts of the $^6$Li
isotope with  isotopic ratios as low as $^7{\rm Li}/^6{\rm Li}=5$. However,
trials using a $^7{\rm Li} / ^6{\rm Li}$ isotopic ratio of 5 and 2 were
performed (see Fig.\ 11), but they gave  a slightly worse numerical fit to the
observed spectrum than the case of pure $^7$Li. Higher S/N  spectroscopic observations will be needed to test
the spallation scenario. Recently, Li has been detected in the companion of the
millisecond pulsar J1740$-$5340 in the globular cluster NGC 6397 (Sabbi et al.
2003). The secondary is a turn-off star which has lost most of its mass and
thus most of its initial lithium content. The high Li abundance measured in
this star suggests that actually some Li production may take place in these
systems. 

\section{Conclusions}

We have obtained a high quality spectrum  of the secondary star in \mbox{A0620$-$00} and 
derived atmospheric chemical abundances. We have set up a technique that
provides a determination of the stellar parameters taking into
\mbox{account} any possible veiling from the accretion disk. We find $T_{\mathrm{eff}} = 4900
\pm 150$ K, $\log g = 4.2 \pm 0.3$, and a veiling (defined as $F_{\rm
disk}/F_{\rm cont,star}$) of less than 15 per cent at 5000 {\AA}  and 
decreasing towards longer wavelengths. Assuming a mass for the secondary
of $M_2 = 0.68 \pm 0.18$ {$M_\odot$}, the  estimated  surface gravity leads to a
stellar radius of $R_2 = 1.1 \pm 0.4$ {$R_\odot\; $},  consistent
with the size of the Roche lobe for the secondary given by Gelino et al. (2001). 
 
The abundances of Fe, Ca, Ti, Al, and Ni are slightly
higher than solar. The abundance ratios of each element with respect to Fe were 
compared with these ratios in late-type main sequence metal-rich stars.
Moderate anomalies for Ti, Ni, and especially Al have been found. A
comparison with element yields from spherically symmetric supernova explosion
models suggests that the secondary star captured part of the ejecta from a
supernova that also originated the compact object in \mbox{A0620$-$00}. The  
abundances can be explained if a progenitor with a $\sim 14$
{$M_\odot\; $} helium core exploded with a mass cut in the range 11--12.5 
{$M_\odot$}, such that no significant amount of iron could escape from the collapse
of the inner layers. Elements such as O, Mg, Si, S, and C, with unavailable 
transitions in our spectrum, will be studied to confirm this scenario.

The Li abundance in the secondary star in \mbox{A0620$-$00} is dramatically enhanced
in comparison with field  late-type main sequence stars, possibly 
indicating either that this is a young system ($\sim0.5$--$2\times10^8$ yr), 
or the existence of
a Li production-preservation mechanism, such as the $\alpha$--$\alpha$ reactions,
which have to be tested analyzing the $^7{\rm Li} / ^6{\rm Li}$ isotopic ratio
using future higher S/N optical spectroscopic observations.  

\section{Acknowledgements}

We are grateful to Hideyuki Umeda and Ken'ichi Nomoto for sending us their
explosion models and several programs for our model computations,
and for helpful discussions. We would also like to thank Jos{\'e} Acosta Pulido 
for his help in the stellar and accretion disk parameter program for searching
the best stellar atmospheric model. We also thank the referee for useful
comments. This work has made use of the VALD database and
IRAF facilities and have  been partially financed by the Spanish Ministry project
AYA2001-1657.

\end{document}